\documentclass{aa}  

\graphicspath{{Images/}} 
\usepackage{graphicx}
\usepackage{subcaption}
\usepackage{txfonts,textcomp}
\usepackage{natbib}
\bibpunct{(}{)}{;}{a}{}{,} 
\begin{document} 

   \title{A deep learning approach for focal-plane wavefront sensing \\ using vortex phase diversity}

   \author{M. Quesnel\inst{1,2}
          \and
          G. Orban de Xivry\inst{2}
          \and
          G. Louppe\inst{1}
          \and
          O. Absil\inst{2}\fnmsep\thanks{F.R.S.-FNRS Senior Research Associate}
          }

   \institute{Montefiore Institute of Electrical Engineering and Computer Science, University of Liège
            \and
             Space sciences, Technologies and Astrophysics Research (STAR) Institute, University of Liège
             }

   \date{Received December 24, 2021; accepted September 30, 2022}


 
  \abstract
   {The performance of high-contrast imaging instruments is limited by wavefront errors, in particular by non-common path aberrations (NCPAs). Focal-plane wavefront sensing (FPWFS) is appropriate to handle NCPAs because it measures the aberration where it matters the most, that is to say at the science focal plane. Phase retrieval from focal-plane images results, nonetheless, in a sign ambiguity for even modes of the pupil-plane phase.}
   {The phase diversity methods currently used to solve the sign ambiguity tend to reduce the science duty cycle, that is, the fraction of observing time dedicated to science. In this work, we explore how we can combine the phase diversity provided by a vortex coronagraph with modern deep learning techniques to perform efficient FPWFS without losing observing time.}
   {We applied the state-of-the-art convolutional neural network EfficientNet-B4 to infer phase aberrations from simulated focal-plane images. The two cases of scalar and vector vortex coronagraphs (SVC and VVC) were considered using a single post-coronagraphic point spread function (PSF) or two PSFs obtained by splitting the circular polarization states, respectively.}
   {The sign ambiguity has been properly lifted in both cases even at low signal-to-noise ratios (S/Ns). Using either the SVC or the VVC, we have reached a very similar performance compared to using phase diversity with a defocused PSF, except for high levels of aberrations where the SVC slightly underperforms compared to the other approaches. The models finally show great robustness when trained on data with a wide range of wavefront errors and noise levels.}
   {The proposed FPWFS technique provides a 100\% science duty cycle for instruments using a vortex coronagraph and does not require any additional hardware in the case of the SVC.}


   \maketitle
%

\section{Introduction}
Because of the small angular separation and high contrast between planetary companions and their parent star, exoplanet imaging is particularly challenging. Although these constraints can be addressed with specific instruments such as coronagraphs, residual wavefront aberrations still represent an inherent obstacle for detecting the majority of exoplanets. To a large extent, these residuals originate from non-common path aberrations (NCPAs) between the scientific and wavefront sensing arms. Focal-plane wavefront sensing (FPWFS) is an approach that has the advantage of taking NCPAs into account by probing their signature in the focal-plane images \citep{Jovanovic:18} while offering high sensitivity.

Estimating phase aberrations from the sole scientific images is not trivial since the relationship between focal-plane intensities and the pupil-plane phase is nonlinear and degenerate \citep{Guyon:18}. Numerical methods have been developed for FPWFS, such as iterative algorithms \citep{Fienup:82}, with the most standard one being the Gerchberg-Saxton algorithm \citep{Gerchberg:72}. More recent techniques have been proposed for various applications \citep[see][for a review]{Jovanovic:18}, including the use of deep learning techniques for FPWFS  \citep{Paine:18, Andersen:19, Andersen:20, Orban:21}. All of these approaches have to deal with one important hindrance: for a centrosymmetric pupil, two different phase distributions in the input pupil plane can produce the same point spread function (PSF). This ambiguity, also called the twin-image problem \citep[e.g.,][]{Guizar-Sicairos:12}, is typically solved with phase diversity using, for instance, an additional defocused PSF \citep{Gonsalves:82}, or an asymmetric pupil mask \citep{Martinache:13}. This, however, reduces the science duty cycle because some observing time, and/or part of the science beam, has to be dedicated to wavefront measurements exclusively. 

Based on the properties of the vector vortex coronagraph \citep[VVC,][]{Mawet:05}, a Nijboer-Zernike phase retrieval approach tailored to the post-VVC PSF was formulated in \citet{Riaud:12b, Riaud:12a}. They proposed to split the two circular polarization states to exploit the phase diversity introduced by the two opposite topological charges associated with the VVC. A similar approach was more recently used by \citet{Bos:19} in the case of the grating-vector apodizing phase plate; although, it also required an asymmetric pupil to lift the sign ambiguity fully.

Here, we revisit the problem of phase retrieval behind a vortex coronagraph using deep learning techniques. Unlike an analytical approach, which could show limitations regarding its formulation, deep learning models can be trained regardless of the instruments and observing conditions. First, in Sect.~\ref{sec:vfpwfs}, we argue that a scalar vortex coronagraph (SVC) has the potential to yield comparable residual phase errors to the dual-polarization VVC implementation, using a single post-coronagraphic PSF instead of two. In Sect.~\ref{sec:deep_learning}, we present our deep learning approach, based on convolutional neural networks (CNNs), which have the advantage of being flexible and easy to implement, and they have already been shown to be capable of reaching fundamental noise limits in our previous works \citep{Quesnel:20,Orban:21}. Finally, in Sect.~\ref{sec:results}, we provide quantitative results on simulated data. We compare the performance of our vortex phase diversity method to a classical approach, and assess the robustness of the models, notably in the presence of representative atmospheric turbulence residuals.

\section{Vortex phase diversity}
\label{sec:vfpwfs}
\subsection{Vortex coronagraphs}
The vortex coronagraph (VC), introduced by \citet{Mawet:05}, is a transparent focal plane mask that diffracts on-axis light outside of the pupil area. A Lyot stop placed in a downstream pupil plane allows this diffracted light to be blocked, enabling high contrast observations. Because of wavefront aberrations, some incoming light from the star is, however, not blocked. Indeed, the VC only removes the Airy disk, and speckles still appear in the focal plane. 

There are two different types of vortex coronagraphs: vectorial (VVC) and scalar (SVC). The VVC applies a geometrical phase ramp to the incoming wavefront with a transmission $t = \exp(\pm j\,l_p\,\theta)$, where $l_p$ is the topological charge and $\theta$ is the azimuthal coordinate. Conjugated phase ramps are applied to each circular polarization state, producing a different signature in the focal plane for each \citep{Riaud:12b}. Here we focus on a topological charge $l_p=2$, which is the most commonly used design so far \citep{Mawet:09, Absil:16}, but the following developments would also hold for any even topological charge. Unlike the VVC, the SVC uses longitudinal phase delays \citep{Ruane:19, Desai:21}, and thereby applies the same phase ramp (e.g., with $+l_p$) to both polarization states. The focal-plane signature behind an SVC corresponds to the one obtained with a single polarization state using the VVC.

\subsection{Sign ambiguity and phase diversity}
In FPWFS, the Fourier relationship between the PSF and the pupil-plane phase causes a sign ambiguity for Zernike modes of an even radial order (e.g., defocus, astigmatism): 
\begin{equation}
    \label{eq:FT}
|\mathcal{F}(E_{\rm even}(x))|^2 = |\mathcal{F}(E_{\rm even}^{*}(-x))|^2,
\end{equation}
where $E_{\rm even}(x) = \text{exp}(-j\phi_{\rm even}(x))$ is the pupil-plane electric field with phase aberrations $\phi_{\rm even}$ (containing even modes only), $E_{\rm even}^{*}$ is its conjugate, and $\mathcal{F}(.)$ is the Fourier transform operator. This sign ambiguity is a strong limitation for FPWFS using a single in-focus image. A fair number of FPWFS methods have been developed to solve the twin-image problem. The most standard one is to use an additional known defocus together with the in-focus image. An illustration of this ambiguity can be found in Fig.~\ref{fig:ex_vvc}, where we generated two phase maps with opposite signs for their even Zernike modes. After propagation through a VVC, the in-focus PSFs are the same in both cases (Fig.~\ref{sfig:psf-d} and \subref{sfig:psf-f}), showcasing the twin-image problem. The out-of-focus PSFs, however, are different (Fig.~\ref{sfig:psf-e} and \subref{sfig:psf-g}) because the added defocus has the same sign in both cases, which allows the ambiguity to be lifted. 

Now, if the two orthogonal circular polarization states are split downstream of the VVC to separate the conjugated phase ramps ($-l_p$ and $+l_p$), or if the case of the SVC is considered, the in-focus PSFs are not identical anymore (Fig.~\ref{sfig:psf-h} and \subref{sfig:psf-j}, or Fig.~\ref{sfig:psf-i} and \subref{sfig:psf-k}). The resulting PSFs are actually switched between the two circular polarization states. This indicates that the sign ambiguity can potentially be lifted when using either the two PSFs obtained from the separate circular polarization states, or the single PSF behind the SVC independently of the polarization state. This illustrates the fact the VC provides an azimuthal phase diversity, which can be used instead of the radial phase diversity provided by an additional defocus \citep{Riaud:12b}. In the case of the SVC, the sign ambiguity would then be lifted similarly to using only an out-of-focus PSF in classical phase diversity \citep[e.g.,][]{Lamb:21}.

\newcommand{\datasize}{0.09}
\begin{figure}[t]
    \centering
    \begin{subfigure}{.99\linewidth}
        \centering
        \includegraphics[scale=\datasize]{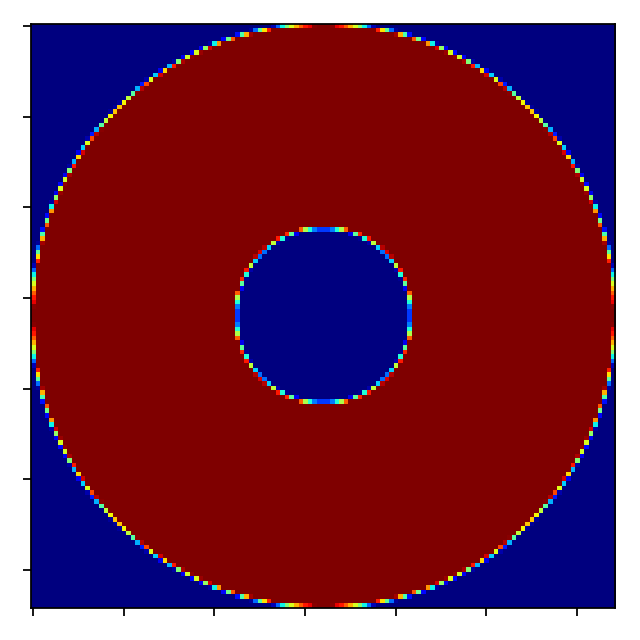}
        \vspace{-5pt}
        \caption{\tiny Entrance pupil}
        \label{sfig:pupil}
    \end{subfigure}\\
    \begin{subfigure}{.49\linewidth}
        \centering
        \includegraphics[scale=\datasize]{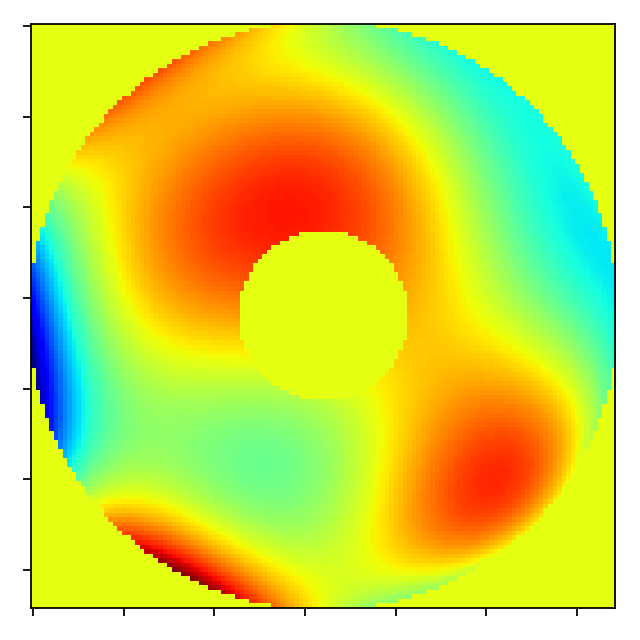}
        \vspace{-5pt}
        \caption{\tiny Phase map $\phi$}
        \label{sfig:phase1}
    \end{subfigure}
    \begin{subfigure}{.49\linewidth}
        \centering
            \includegraphics[scale=\datasize]{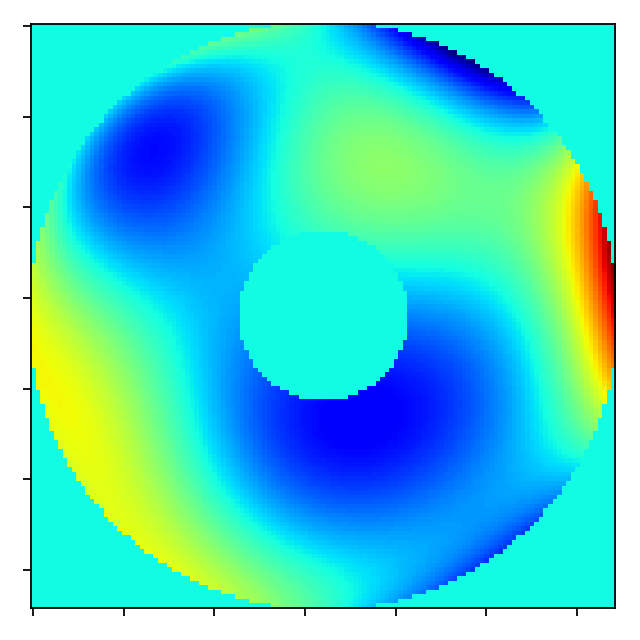}
        \vspace{-5pt}
        \caption{\tiny Phase map $\phi' (\phi'_{even} =  - \phi_{even})$}
        \label{sfig:phase2}
    \end{subfigure}\\
    \begin{subfigure}{.22\linewidth}
            \includegraphics[scale=\datasize]{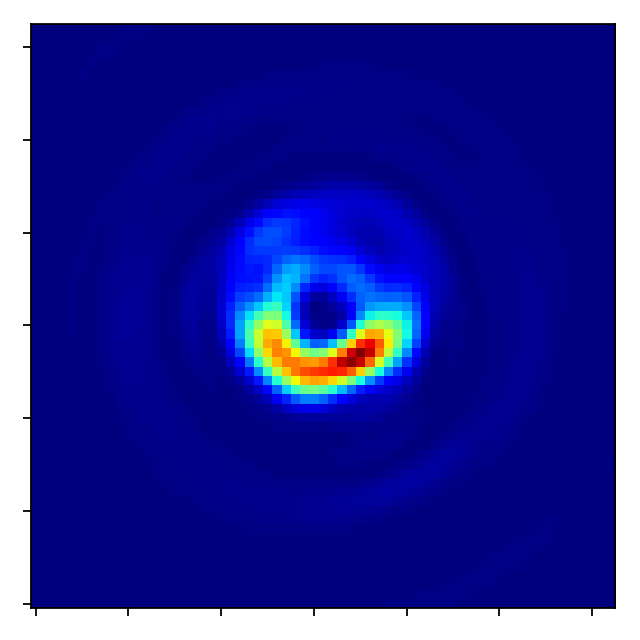}
        \vspace{-15pt}
        \caption{\tiny $\pm l_p$}
        \label{sfig:psf-d}
    \end{subfigure}
    \begin{subfigure}{.22\linewidth}
        \includegraphics[scale=\datasize]{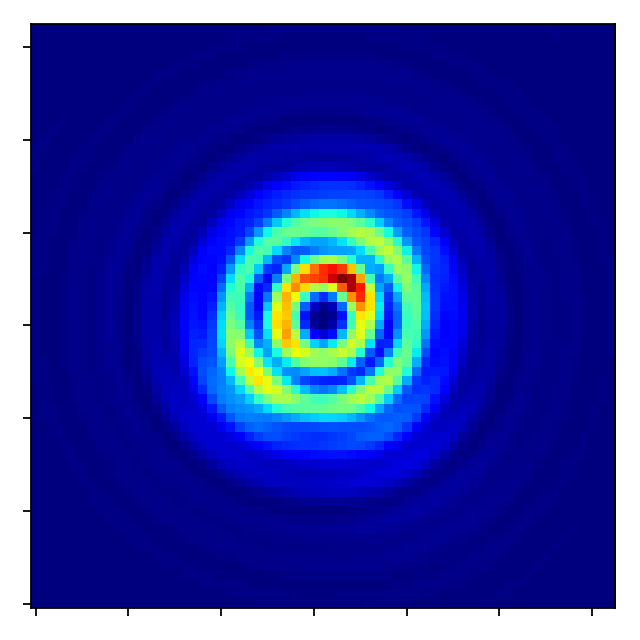}
        \vspace{-15pt}
        \caption{\tiny $\pm l_p$; out}
        \label{sfig:psf-e}
    \end{subfigure}
    \hspace{10pt}
    \begin{subfigure}{.22\linewidth}
            \includegraphics[scale=\datasize]{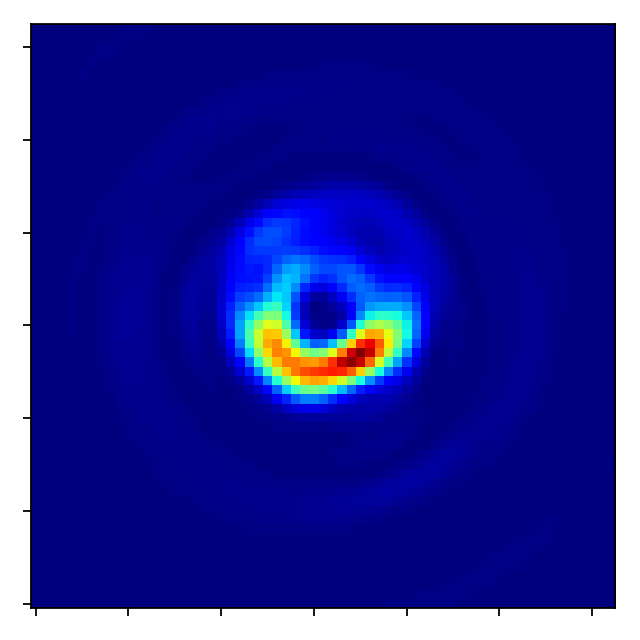}
        \vspace{-15pt}
        \caption{\tiny $\pm l_p$}
        \label{sfig:psf-f}
    \end{subfigure}
    \begin{subfigure}{.22\linewidth}
            \includegraphics[scale=\datasize]{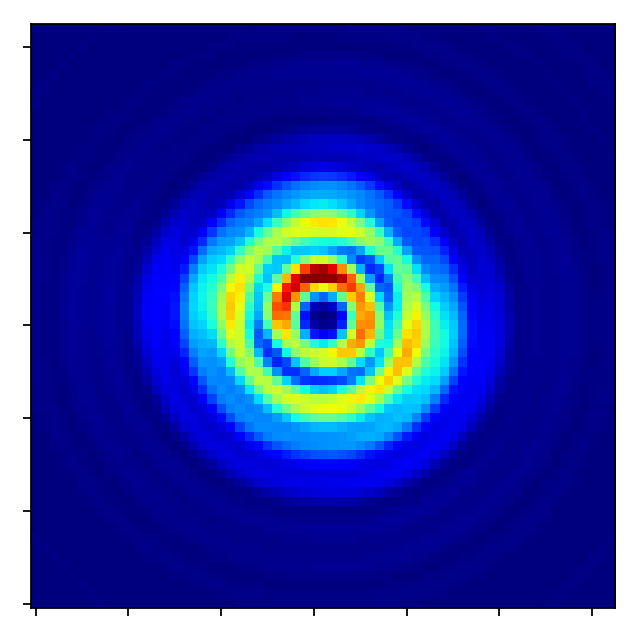}
        \vspace{-15pt}
        \caption{\tiny $\pm l_p$; out}
        \label{sfig:psf-g}
    \end{subfigure}\\
    \begin{subfigure}{.22\linewidth}
            \includegraphics[scale=\datasize]{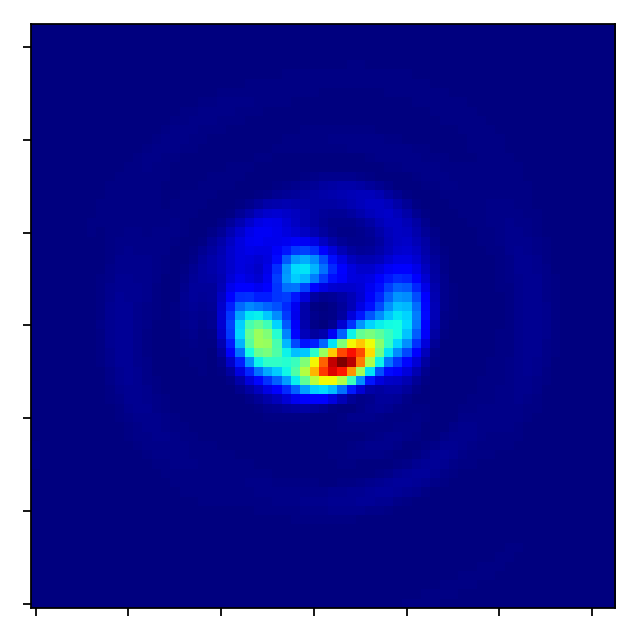}
        \vspace{-15pt}
        \caption{\tiny $+l_p$}
        \label{sfig:psf-h}
    \end{subfigure}
    \begin{subfigure}{.22\linewidth}
            \includegraphics[scale=\datasize]{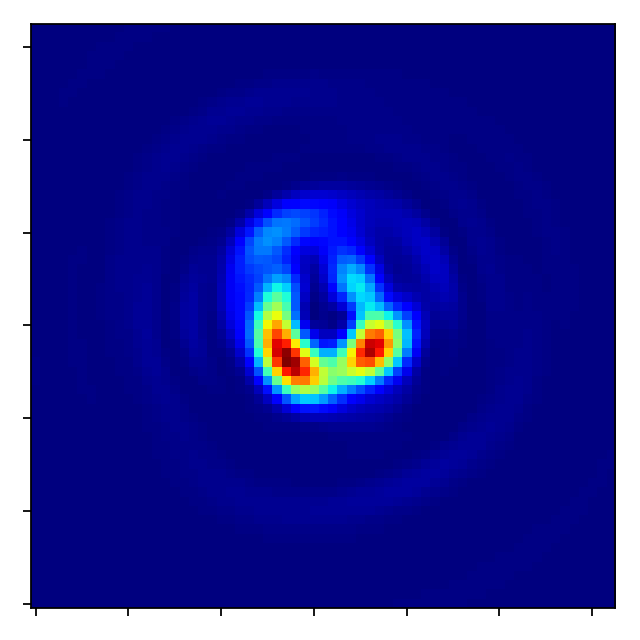}
        \vspace{-15pt}
        \caption{\tiny $-l_p$}
        \label{sfig:psf-i}
    \end{subfigure}
    \hspace{10pt}
    \begin{subfigure}{.22\linewidth}
            \includegraphics[scale=\datasize]{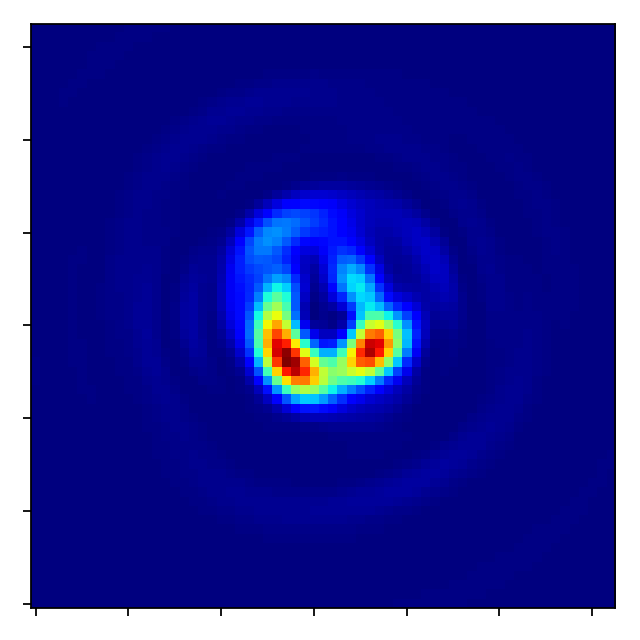}
            \vspace{-15pt}
        \caption{\tiny $+l_p$}
        \label{sfig:psf-j}
    \end{subfigure}
    \begin{subfigure}{.22\linewidth}
            \includegraphics[scale=\datasize]{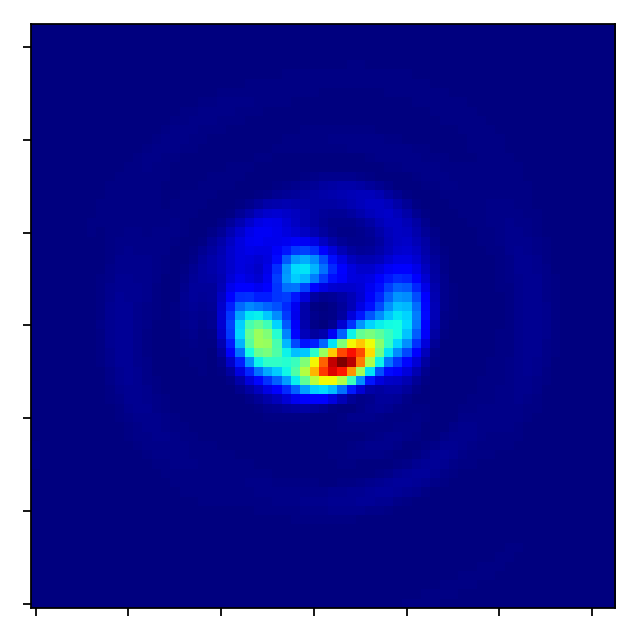}
            \vspace{-15pt}
        \caption{\tiny $-l_p$}
        \label{sfig:psf-k}
    \end{subfigure}
    \caption{Comparison of simulated PSFs for two conjugated phase maps $\phi$ (\textit{left}) and $\phi'$ (\textit{right}): for $\phi'$, we set opposite Zernike coefficients to those of $\phi$ only for the even modes, with a total of 18 modes starting from defocus. \textbf{(a):}~Entrance annular pupil. \textbf{(b,~c):}~The conjugated phase maps. \textbf{(d,~e):}~In-focus and out-of-focus PSFs obtained from propagating (b) with both polarization states together. \textbf{(f,~g):}~The same as (d,~e) but using (c) for propagation instead. \textbf{(h,~i):} In-focus PSFs obtained from (b) with $-l_p$ and $+l_p$ used separately. \textbf{(j,~k):}~The same as (h, i) but using (c) for propagation instead.}
    \label{fig:ex_vvc}
\end{figure}

\section{Deep learning approach}
\label{sec:deep_learning}

\subsection{Data generation}
\label{sec:data_gen}

In our simulations, we considered an annular entrance pupil with a diameter of 8~m and a central obstruction of 30\%. An observed bandwidth of 0.2~\textmu m was defined around 2.2~\textmu m (K band), by simulating a total of five wavelengths. A pixel scale of $0.25\,\lambda / D /$pix was set with a detector containing 64$\times$64 pixels, giving a field-of-view of $16\,\lambda/D$. The most relevant simulation parameters are listed in Table~\ref{tab:data_param}. 

We generated the phase aberrations using annular Zernike polynomials, which make up an orthonormal basis on the input pupil:
\begin{equation}
    \label{eq:phase}
    \phi(x, y) = \sum_{i=1}^{N_{\rm modes}} c_i \, Z_i(x, y),
\end{equation}
where $\phi$ is the complete phase map, $Z_i$ are the Zernike polynomials, $c_i$ are the corresponding coefficients, and $N_{\rm modes}$ is the number of modes considered.

The generated datasets are composed of 18 or 88 Zernike modes, up to the fifth and 12th radial orders, respectively, excluding the piston, tip, and tilt modes. The set of Zernike coefficients for each sample was first randomly generated within the range $[-1,1]$ before each coefficient was divided by its corresponding radial order to approximate a $1/f^2$ power spectral density profile, typically encountered with good quality optics \citep{Dohlen:11}. Low and high aberration levels, represented by wavefront error (WFE) distributions centered at a 70 and 350~nm root mean square (RMS), respectively, are considered by normalizing the Zernike coefficients accordingly. An example of such a distribution can be seen in \citet{Orban:21}. For classical phase diversity, the additional defocus was set to $\lambda$/5, that is, 440~nm RMS. In our case, this amount of diversity is close to the optimal value in terms of phase retrieval performance. The defocus was added in the entrance pupil plane, as if done by the deformable mirror of an adaptive optics system, which means that the resulting defocused PSFs contain more flux than the in-focus PSFs as the coronagraphic performance of the VC is degraded.

To increase the representativeness of our simulations and to test the robustness of our approach, we added atmospheric turbulence residuals to the phase maps. A state-of-the-art extreme adaptive optics (AO) was simulated using the COMPASS library \citep{Ferreira:18}, assuming a loop frequency of 3.5~kHz, 2-frame delay, a 50$\times$50 deformable mirror (i.e., 2040 modes/valid actuators), and a pyramid sensor with $5~\lambda$/D of modulation (without noise). This has yielded a Strehl ratio of about 98\% at 2.2~\textmu m, corresponding to a WFE of about 50~nm RMS. We sampled the AO residuals at 10~Hz and we used a sequence of ten consecutive phase screens by summing up the corresponding PSFs. We therefore simulated a 1-s exposure in the presence of a given amount of static NCPAs. The results with data containing these AO residuals are shown in Sect.~\ref{sec:results_robust}.

To simulate a PSF obtained behind a VVC, we performed two propagations, one with $+l_p$ and the other with $-l_p$, to consider each circular polarization state. The downstream Lyot stop blocked 2\% of the outer pupil area (but the central obstruction was not oversized). The resulting PSFs were then either summed up to reproduce the nonpolarized case, or they were kept separate to consider the dual-polarization case. To simulate the SVC, only one such PSF was taken. The optical propagation was handled by the HEEPS package\footnote{\url{https://github.com/vortex-exoplanet/HEEPS}} \citep{Carlomagno:20}, which makes use of PROPER \citep{Krist:07}. Examples of generated phase maps and PSFs can be found in Fig.~\ref{fig:ex_vvc}. We then added photon noise to our PSFs, so that the signal-to-noise ratio (S/N) was defined as $ S/N = \sqrt{N_{\rm ph}}$, where $N_{\rm ph}$ is the number of photons. A square-root stretching operation was applied to the PSFs to help the CNN identify the speckle patterns. Finally, we normalized the PSFs with a min-max scaling to obtain flux in the range [0,1], which ensured the CNN was fed with same-scale quantities.

\begin{table}[t]
    \caption[]{Data generation parameters}
    \label{tab:data_param}
    
    \centering
    \begin{tabular}{cc}
    \hline \hline
        Parameter      &  \text{Value} \\
        \hline
        Central obstruction     & 30 \%            \\
        Topological charge     & 2            \\
        Pixel scale & $0.25\,\lambda / D$ / px     \\
        Field-of-view           & $16\,\lambda/D$  \\
        \noalign{\smallskip}
        \hline
    \end{tabular}
\end{table}

\subsection{Model architecture}

We built deep neural network models whose goal is to map the Zernike coefficients of phase aberrations $\phi$ from a given PSF $I$, that is, to approximate a nonlinear function $f$ such that $\phi \approx f(I)$. CNNs have been proven to be very well suited for image analysis, with numerous applications for both classification and regression tasks. CNN-based architectures have been developing very quickly in recent years, with performance still improving greatly. We have therefore used a state-of-the-art deep CNN called EfficientNet \citep{Tan:19}. This type of architecture stands out from other ones by using a new scaling technique: all dimensions of the CNN (depth, width, and resolution) are scaled by the same compound coefficient $\Phi$, and the parameters are inferred from the original model or baseline EfficientNet-B0 ($\Phi = 0$). There are thus different models available, and we chose to use EfficientNet-B4, for which we have obtained the best trade-off between model performance and runtime. EfficientNet-B4 has a total of $1.9 \times 10^{7}$ parameters and 4.2 $\times 10^{9}$ FLOPS. It has about the same number of parameters as the ResNet-50 architecture, which was used in \citet{Quesnel:20} and \citet{Orban:21}.

\subsection{Model training}

For a given training, a dataset composed of $10^{5}$ PSFs (or PSF pairs for the cases with two input channels) was randomly split into training (90\%) and validation (10\%) sets. Each sample also contains the true NCPA phase maps as labels, while the AO phase screens are never given. Batches composed of 64 data samples were then consecutively fed to the neural network. We define the loss function as the root-mean-square error (RMSE) of the phase residuals. Weight updates based on the loss were handled by the Adam optimizer \citep{Kingma:17}. To improve the performance, we set a penalty on the loss (``weight decay'') of $10^{-7}$ for the low aberration regime and $10^{-6}$ for the higher aberration regime. We also set an initial learning rate of $10^{-3}$ which was decreased by a factor of two as soon as the validation loss reached a plateau over 15 epochs. This results in sudden loss drops, allowing the performance to be greatly improved. Pre-trained models on ImageNet were used to initialize the weights. The training of the model was stopped if no improvement of the validation loss was observed over 25 epochs. This results in training procedures lasting between 50 and 250 epochs.

\section{Results}
\label{sec:results}

\begin{figure*}[t]
    \begin{subfigure}{.49\linewidth}
        \centering
        \includegraphics[width=\hsize]{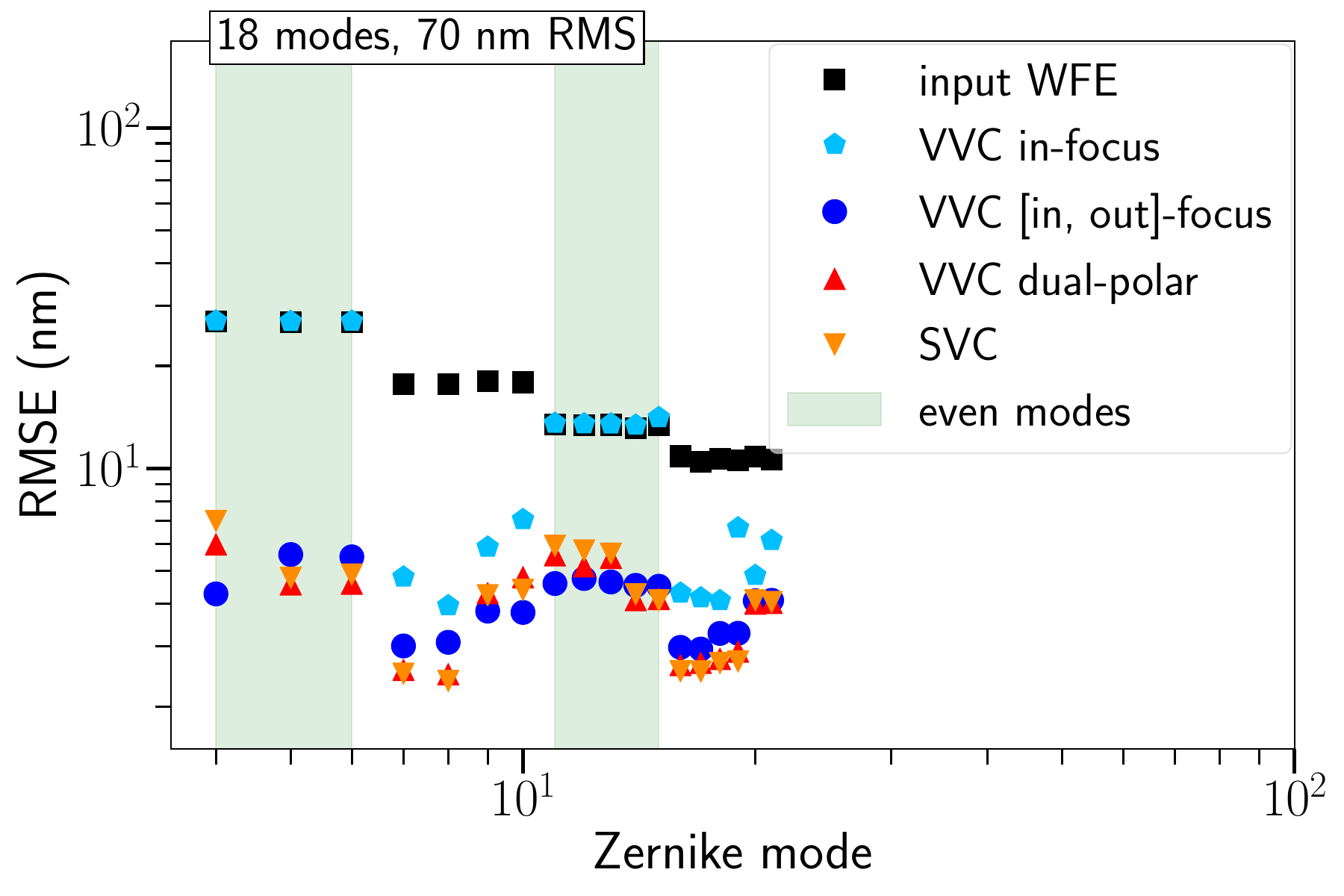}
    \end{subfigure}
    \begin{subfigure}{.49\linewidth}
        \centering
        \includegraphics[width=\hsize]{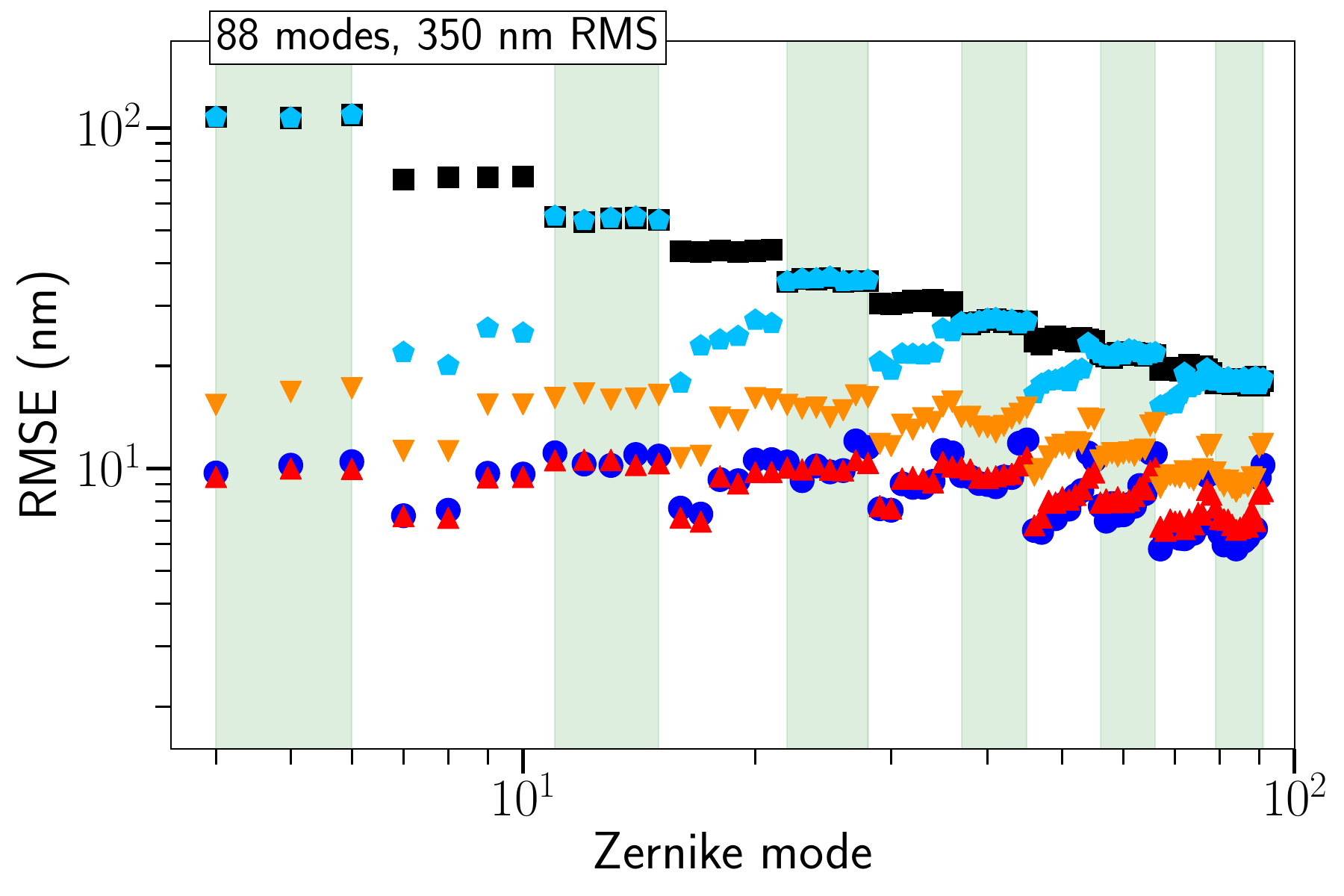}
    \end{subfigure}
        \caption{RMSE per Zernike mode, following the Noll convention, starting from the defocus mode. Four cases were compared (see Table~\ref{tab:type_models} for notations), using a single in-focus post-VVC PSF without splitting the polarization states (cyan), two post-VVC PSFs with additional defocus (dark blue), the two post-VVC PSFs associated with each polarization state (red), and a single PSF after the SVC (orange). The RMSE of the input phase maps is represented in black and the even modes are indicated by the green areas. \textit{Left}: Input WFE of 70~nm distributed over 18 modes. \textit{Right}: Input WFE of 350~nm distributed over 88 modes. In both examples, the S/N in the entrance pupil plane is equal to 100.}
    \label{fig:sign_ambi}
\end{figure*}

We compare the capacity of different configurations to lift the sign ambiguity as well as their performance. The designation of these configurations, together with some of their parameters, can be found in Table~\ref{tab:type_models}: we consider the cases of the VVC with or without classical phase diversity (``VVC [in, out]-focus'' and ``VVC in-focus,'' respectively), which are compared to the new approaches presented in this paper (``VVC dual-polar'' and ``SVC''). The noncoronagraphic case (``no vortex [in, out]-focus'') is evaluated as well. We also investigate the possibility to work with differential PSFs obtained by subtracting the separate circular polarization states (``VVC dual-polar; diff PSFs''). In the last part of this section, we add atmospheric turbulence residuals and we assess the robustness of the models regarding variations in the S/N levels, input wavefront errors, and Zernike polynomial orders. All models are evaluated using 1000 test samples.

\begin{table}[t]
    \caption[]{Configurations considered for phase retrieval}
    \label{tab:type_models}
    \centering
    \begin{tabular}{cccc}
    \hline \hline
        Designation      & Charge & Defocus & Inputs \\
        \hline
        VVC in-focus & $\pm l_p$ & no & 1 \\
        VVC [in, out]-focus & $\pm l_p$ & yes & 2 \\
        VVC dual-polar & [+$l_p$, -$l_p$] & no & 2 \\
        SVC & +$l_p$ & no & 1 \\
        no vortex [in, out]-focus & 0 & yes & 2\\
        VVC dual-polar; diff PSFs & [+$l_p$, -$l_p$] & no & 1 \\
        \hline
     \end{tabular}
\end{table}

\subsection{Phase sign determination}

To determine whether the models predict the correct sign, we looked at the performance per Zernike mode. The metric used is the RMSE per mode: 
\begin{equation}
    \sigma_{z} = \sqrt{\frac{1}{N_{\rm test}} \, \sum_{i}^{N_{\rm test}} (\hat{c}_i - c_i)^2} \: ,
\end{equation}
where $N_{\rm test}$ is the number of test samples, while $\hat{c}$ and $c$ are the estimated and true Zernike coefficients, respectively.

In Fig.~\ref{fig:sign_ambi} we compare the performance per mode between four cases for two different aberration contents. A network using only in-focus PSFs in the nonpolarized case with the VVC yields no correction for even Zernike modes, because the model tends to predict zero for the coefficients facing the ambiguity (due to the $l_2$-norm training loss). For odd modes, the model is able to provide some correction, even though its quality is limited by the loss function, which does not discriminate between even and odd modes. Adding defocused PSFs as input solves the problem as expected \citep{Quesnel:20}. In the dual-polarization case, a network using either one or both circular polarization states separately as input (SVC and VVC, respectively) also yields good performance for even modes as well as for odd modes. This indicates that the sign ambiguity is properly lifted with these two approaches.

It is noteworthy that the performance marginally depends on the Zernike mode: the error tends to increase for larger angular azimuthal orders at a given radial order. Our interpretation is that since the phase information is of higher spatial frequency and located closer to the edge of the pupil in these cases, it is more difficult for the CNN model to identify those features.

\subsection{Performance compared to classical phase diversity}

\begin{figure*}[t]
    \begin{subfigure}{.49\linewidth}
        \centering
        \includegraphics[width=\hsize]{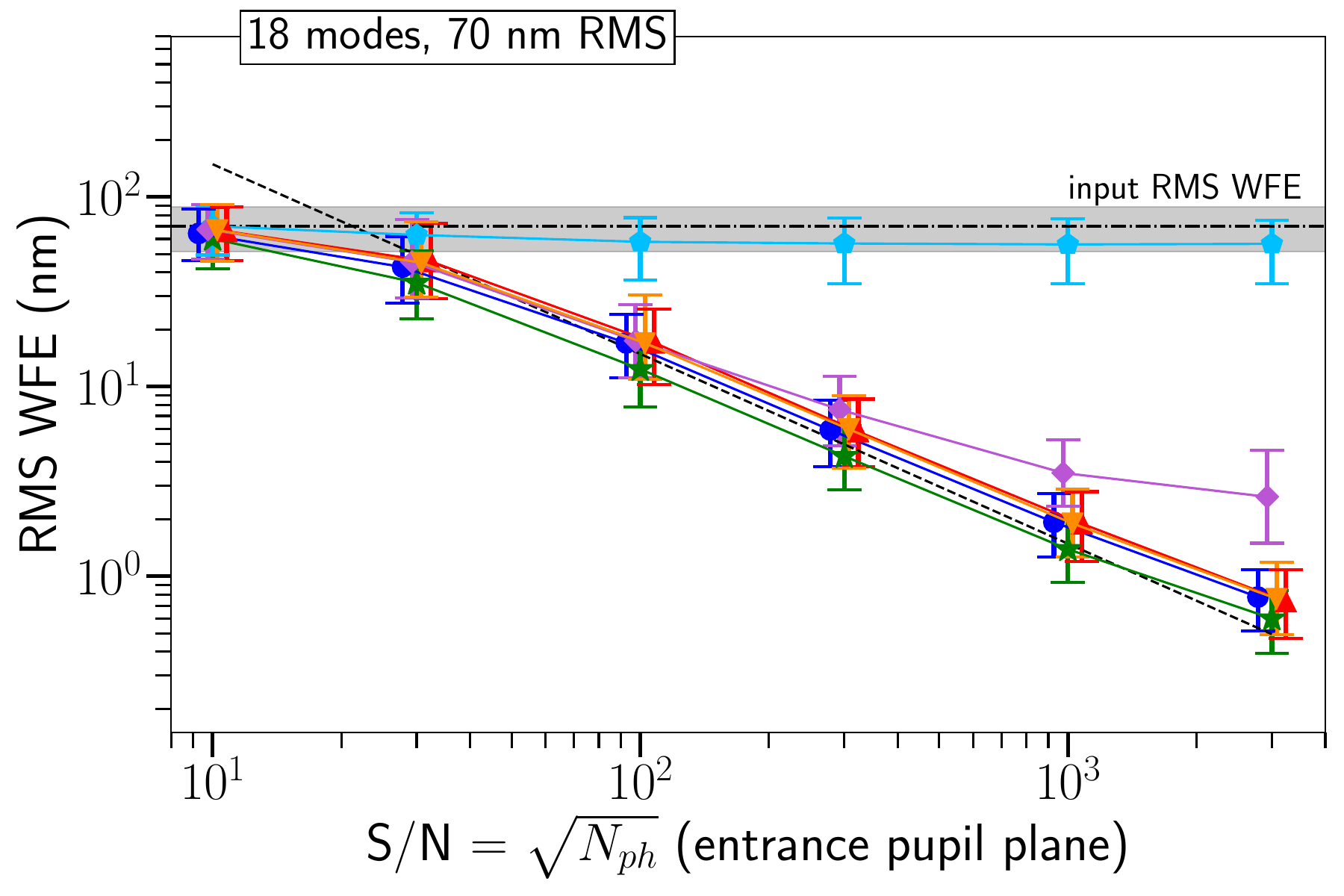}
    \end{subfigure}
    \begin{subfigure}{.49\linewidth}
        \centering
        \includegraphics[width=\hsize]{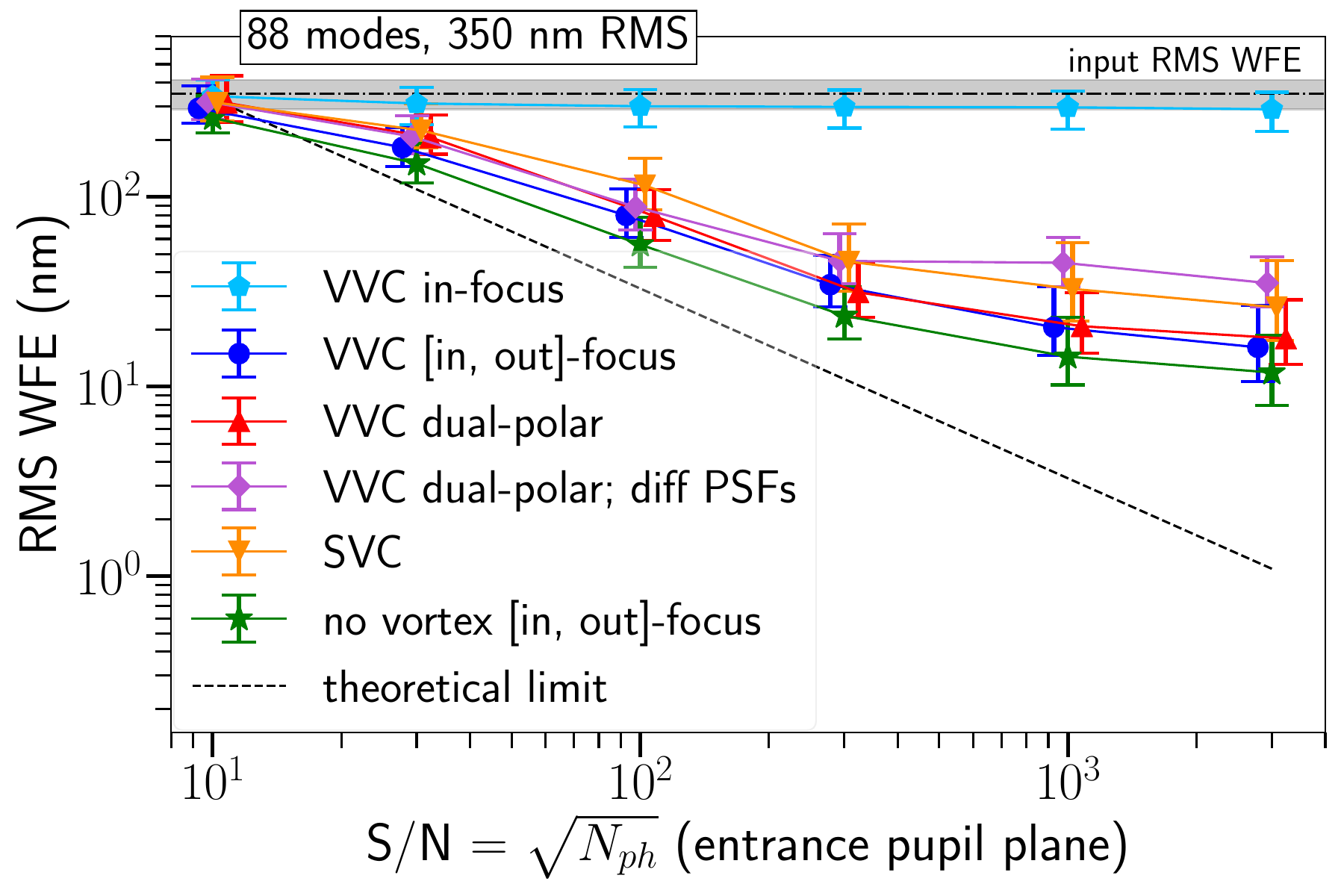}
    \end{subfigure}
    \caption{Performance in terms of RMS WFE on the phase residuals at different S/N levels. Each point corresponds to a model trained and evaluated on the indicated S/N (six S/Ns are considered, and slight horizontal shifts were applied to be able to discern each point). The same colors in Fig.~\ref{fig:sign_ambi} are used, with the addition of the performance with classical imaging (green), using differential post-VVC PSFs (violet), as well as the theoretical limit (black dashed line). The median values are represented and the error bars correspond to the 2-98th percentiles. The S/Ns indicated are the ones at the entrance pupil plane, and the flux suppression induced by the vortex mask is taken into account. \textit{Left}: Input WFE of 70~nm distributed over 18 modes. \textit{Right}: Input WFE of 350~nm distributed over 88 modes.}
    \label{fig:perfo_star}
\end{figure*}

We now compare our method to the classical phase diversity approach in terms of overall phase retrieval performance. The RMS WFE on the phase residuals is used as a metric and it is defined for each test sample as:
\begin{equation}
    \sigma_{\phi} = \sqrt{\frac{1}{N_{\rm pix}}\sum_{i}^{N_{\rm pix}} (\hat{\phi}_i - \phi_i)^2},
\end{equation}
where $N_{\rm pix}$ is the number of pixels, while $\hat{\phi}$ and $\phi$ are the estimated and true pupil phases, respectively. 

In our simulations, we consider the fact that the vortex coronagraphs block out most of the starlight, and that for a given stellar magnitude, the resulting flux in the detector plane is reduced. The flux is also equally split between each PSF for all the cases with two channels, while for the configurations with a single one, the PSF receives the total remaining flux behind the vortex mask. The performance of the trained models at different S/N levels defined in the entrance pupil plane is shown in Fig.~\ref{fig:perfo_star}. In our case, S/Ns between $10^{1}$ and ${3\times10^{3}}$ correspond to stars of apparent magnitudes in the range from 18.6 to 6.2.\footnote{with an integration time of 1s, a transmission and quantum efficiency equal to 50\%, a telescope diameter of 8 m, and a filter bandwidth of 50~nm.} For a median input WFE of 70~nm with 18 modes (Fig.~\ref{fig:perfo_star}, left), the simulated performance is almost identical for the classical, SVC and VVC dual-polarization approaches, even though the additional defocus increases the overall S/N at the focal plane for the classical method. For a median input WFE of 350~nm with 88 modes (Fig.~\ref{fig:perfo_star}, right), the phase residuals are distinctly higher for all the configurations, and a plateau is reached for S/Ns above 1000. We can especially notice that the sole PSF behind the SVC somewhat limits the performance in this case. Our main hypothesis for this discrepancy is that, in a high aberration regime, the effects of the nonlinear nature of the problem are greater. The extra information given by having two input channels is therefore favorable and makes the models easier to train. In general, it is more difficult to train datasets containing strong aberrations, and this can typically be improved by using more data \citep[e.g., $5\times10^5$ samples, see][]{Orban:21}, more complex architectures (e.g., EfficientNet-B6), and/or stronger weight decay.  

We also consider the possible presence of planetary companions in the detected images. This additional, off-axis source of light is largely unaffected by the vortex phase ramp and therefore adds the same signature in both circular polarization states. This additional light source may bias the phase retrieval process, and lead to unwanted planetary signal subtraction. A possible work-around is to subtract one polarization image from the other, in an attempt to remove the signature of any off-axis light source. We thus assessed the phase retrieval capabilities using the difference between both polarization states after the VVC. The results are shown in Fig.~\ref{fig:perfo_star} and are compared with the other configurations. We only obtain a marginal increase in the error at high S/Ns, especially in the low aberration regime, which can be explained by the loss of information produced by subtracting one PSF from the other.

The performance of the various configurations are finally compared to the theoretical limit in Fig.~\ref{fig:perfo_star}. This limit is discussed in \citet{Orban:21} for noncoronagraphic imaging. For both the noncoronagraphic and vortex imaging cases, the residual errors reach the fundamental limit in the low aberration regime (Fig.~\ref{fig:perfo_star}, left). In a higher aberration regime, the performance does not reach the fundamental limit, and the gap increases toward higher S/Ns (Fig.~\ref{fig:perfo_star}, right). This can be improved with more robust training as explained above. One can note that the residual errors are constrained by the WFE distribution in the data toward lower S/Ns, while the theoretical limit is independent of the input WFE distribution and continues to increase for lower S/Ns, thus yielding residual WFE below the limit.

\subsection{Model robustness}
\label{sec:results_robust}

\begin{figure*}[t]
    \begin{subfigure}{.49\linewidth}
        \centering
        \includegraphics[width=\hsize]{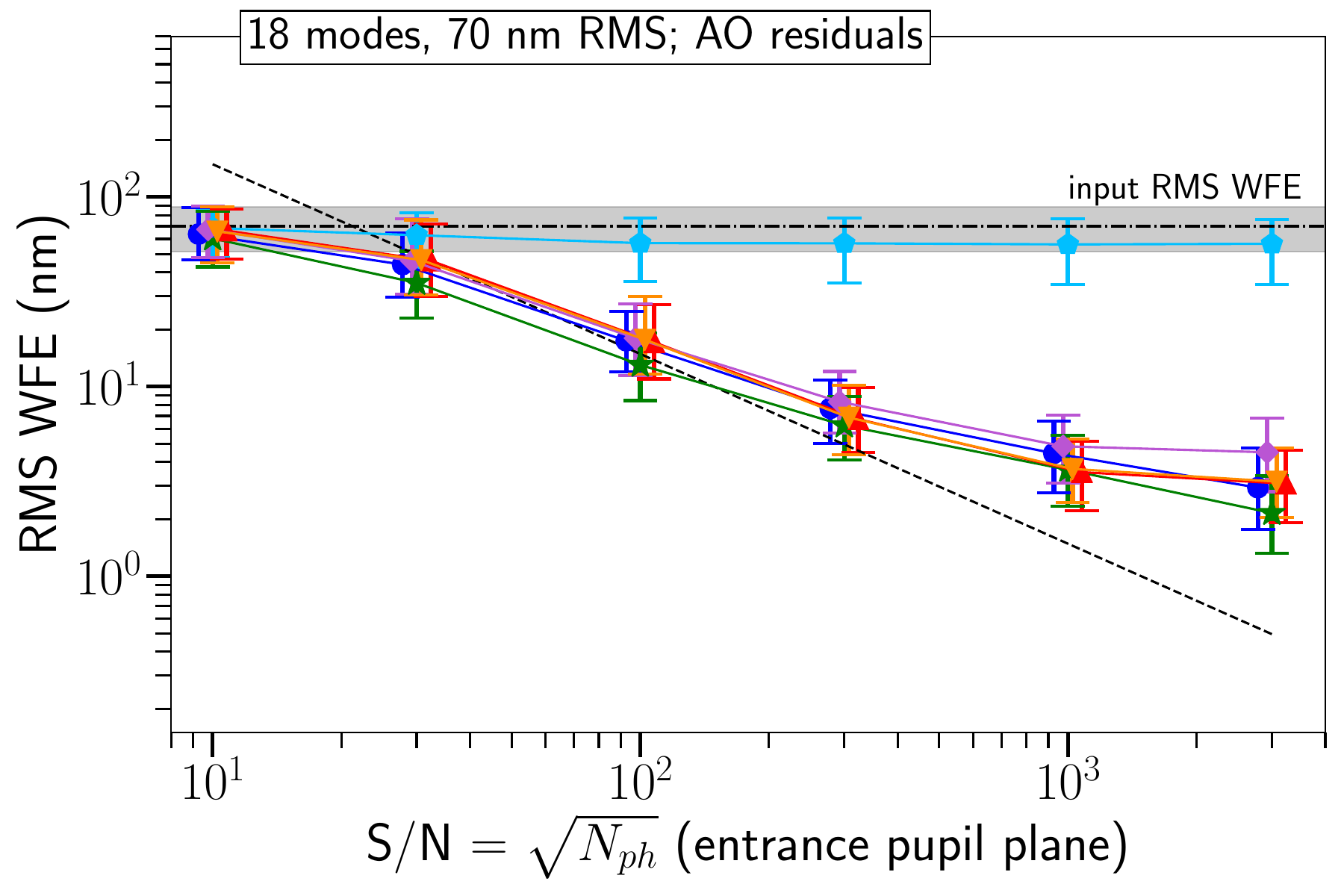}
    \end{subfigure}
    \begin{subfigure}{.49\linewidth}
        \centering
        \includegraphics[width=\hsize]{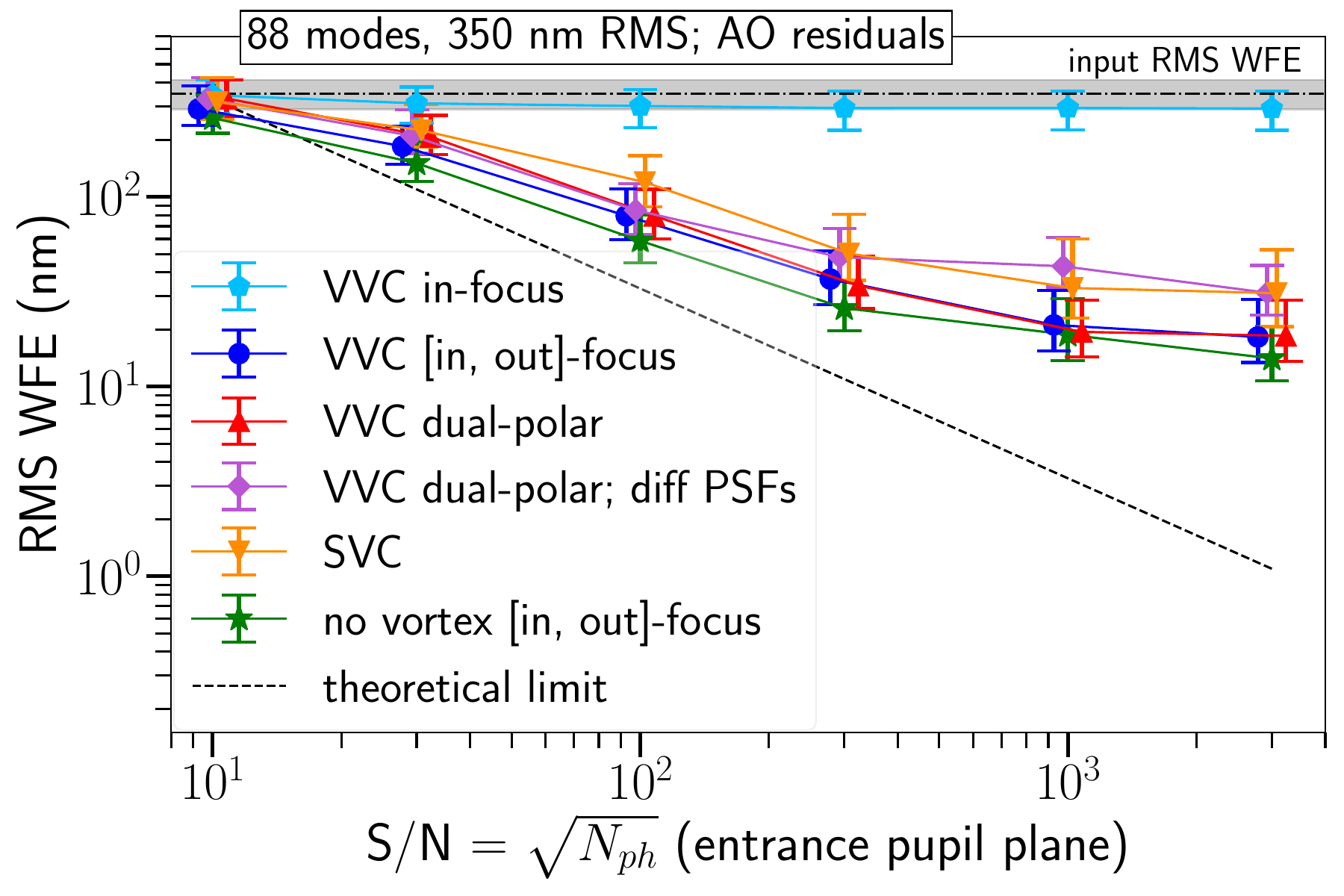}
    \end{subfigure}
    \caption{Phase prediction errors at different S/N levels, presented the same way as in Fig.~\ref{fig:perfo_star}, but this time also including atmospheric turbulence residuals in the PSFs during both training and testing.}
    \label{fig:perfo_star_ao}
\end{figure*}

\begin{figure}[t]
    \begin{subfigure}{.99\linewidth}
        \centering
        \includegraphics[width=\hsize]{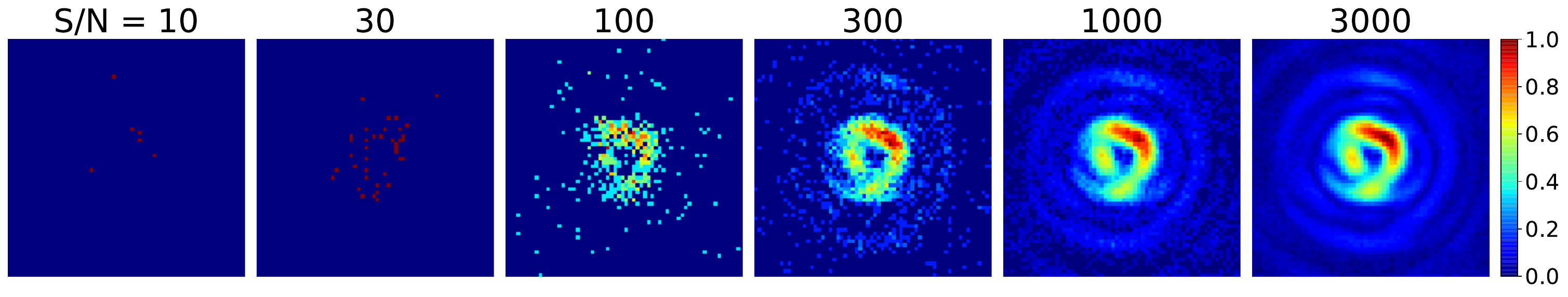}
    \end{subfigure}
    \begin{subfigure}{.99\linewidth}
        \centering
        \includegraphics[width=\hsize]{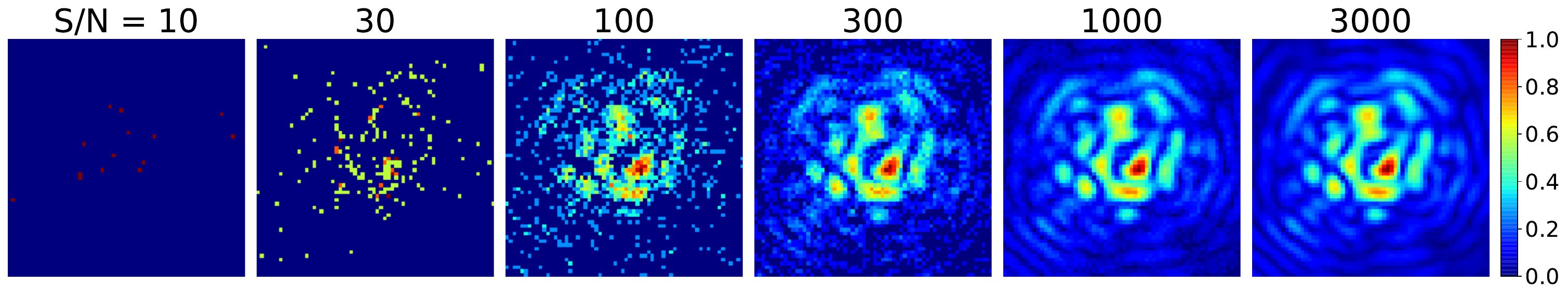}
    \end{subfigure}
    \caption{Examples of PSFs at different S/N levels (defined in the entrance pupil plane) for $+l_p$. The resulting S/N in the detector place is reduced due to the extinction factor introduced by the coronagraph and by the beam splitting between the two polarization channels. The level of NCPA is equal to 70~nm RMS distributed over 18 modes (\textit{top}) and 350~nm RMS over 88 modes (\textit{bottom}). AO residuals are also present: each PSF is the result of combining ten PSFs, with each containing a different AO residual phase screen.}
    \label{fig:psfs_snr}
\end{figure}

\begin{figure}[t]
    \centering
    \includegraphics[width=\hsize]{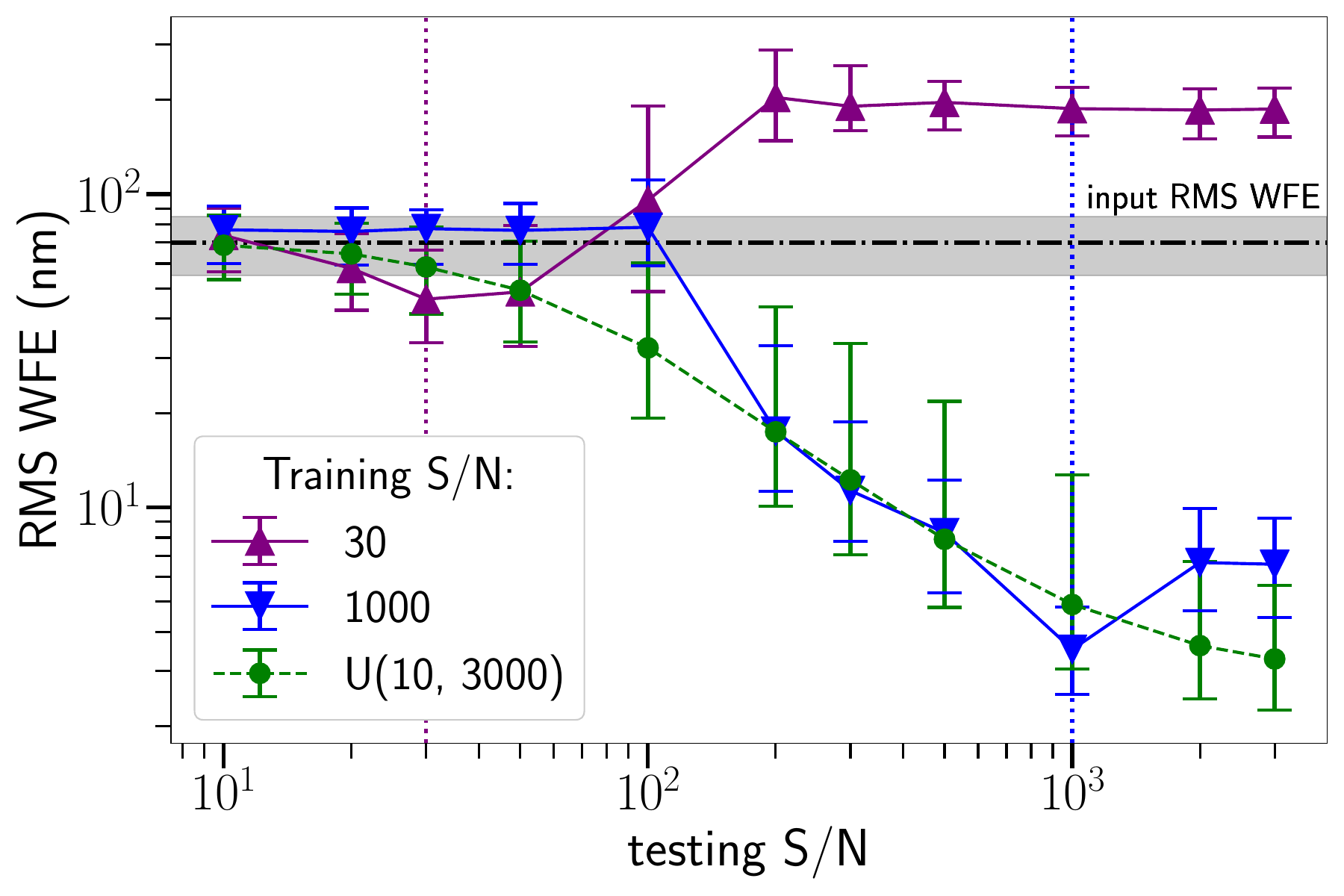}
    \caption{Performance with altered S/N levels during evaluation for three models trained on data with a median RMS WFE of 70~nm over 18 modes, with an S/N of 30 (purple), 1000 (blue), and with S/Ns uniformly distributed over the entire S/N range (green). Each point is obtained from an testing batch composed of 1000 samples (the median value together with the 2-98th percentiles are shown).}
    \label{fig:robust_snr}
\end{figure}

To test how the method handles more realistic ground-based observations, we added atmospheric turbulence residuals in addition to the NCPAs, as described in Sect.~\ref{sec:data_gen}. This represents an additional source of noise since the AO residuals are not included in the labels for training. Examples of input PSFs at the different flux levels can be found in Fig.~\ref{fig:psfs_snr}. The performance now 
starts to reach a plateau of a few nm RMS in the low aberration regime at high S/Ns (Fig.~\ref{fig:perfo_star_ao}, left), due to the presence of these atmospheric turbulence residuals. In the high NCPA regime (Fig.~\ref{fig:perfo_star_ao}, right), the AO residuals however become negligible and the performance is almost identical to the case without turbulence (Fig.~\ref{fig:perfo_star}, right).

We finally study the robustness of the models regarding a variation in the data during evaluation. First, we may encounter different flux levels than those considered during training. In Fig.~\ref{fig:robust_snr}, we illustrate how models in the VVC dual-polar configuration trained on data containing 70~nm RMS behave in such conditions. Whether the training S/N is low or high, models only show good robustness to other flux levels within a limited range, outside of which the performance is strongly degraded. If a more robust model is required, it is also possible to train with various flux levels. We investigated this by using a training dataset covering the entire test S/N range, without increasing its size. The median performance is much more consistent at every S/N; although, the variation in the residual error between samples is greater, and a small degradation can naturally be seen compared to using identical training and testing S/N (as shown in Fig.~\ref{fig:perfo_star_ao}).

\begin{figure}[t]
    \centering
    \includegraphics[width=\hsize]{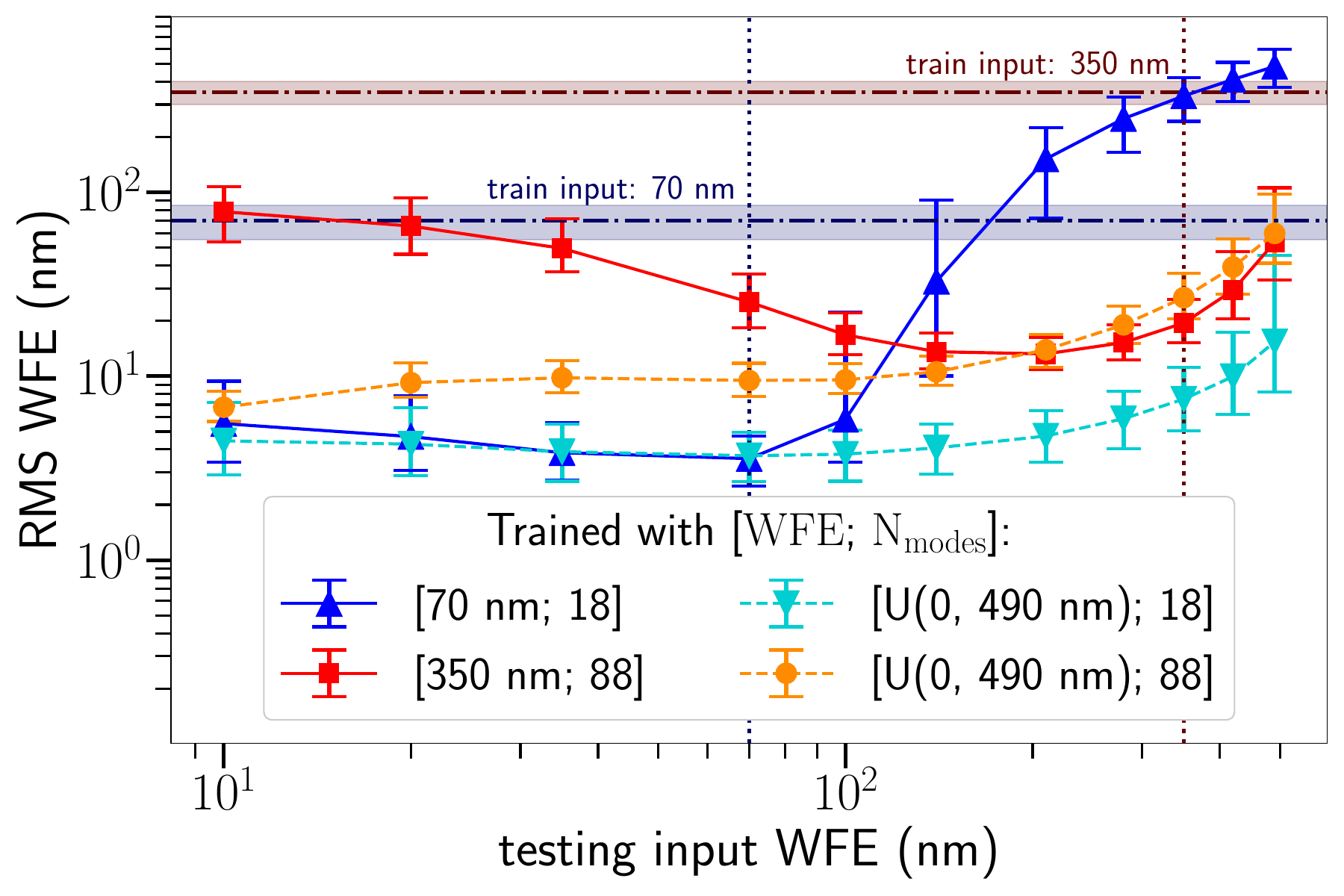}
    \caption{Performance with different input WFE levels defined during evaluation for models trained on data with a median RMS WFE of 70~nm over 18 modes (blue), and 350~nm RMS over 88 modes (red). Models were also trained on data following a uniform distribution covering the whole input WFE range, using both spatial frequency regimes (cyan and orange). The S/N is 1000 and each training dataset contains 10$^5$ samples.}
    \label{fig:robust_wfe}
\end{figure}

We also study the change in performance when evaluating the model outside the input WFE training range. Fig.~\ref{fig:robust_wfe} shows the robustness of models trained on the two aberration regimes studied in this paper. Data containing more aberrations rapidly deteriorate the reconstruction. The models perform better when evaluated at lower aberration levels, but they have limited performance when trained in the high aberration regime. To overcome these limitations, we trained two models over the entire test WFE range for each of the Zernike mode contents considered in the paper. Such models show excellent robustness, with minimal degradation compared to models with identical training and testing WFE distributions. This suggests that these models could be robustly used in closed-loop operations, even with the aberration level decreasing with time. Regarding the varying spatial power spectral density of the wavefront, the residuals are generally constant along the Zernike modes, as seen in Fig.~\ref{fig:sign_ambi}. When giving the reconstructed PSFs as input to the same trained model, we have observed that most residual RMS WFE stay below 10~nm for a model trained on 70~nm RMS as input and an S/N of 1000. A thorough analysis of a closed-loop application will be the subject of future work when testing the algorithm in the lab or on-sky.

\begin{figure}[t]
    \begin{subfigure}{.99\linewidth}
        \centering
        \includegraphics[width=\hsize]{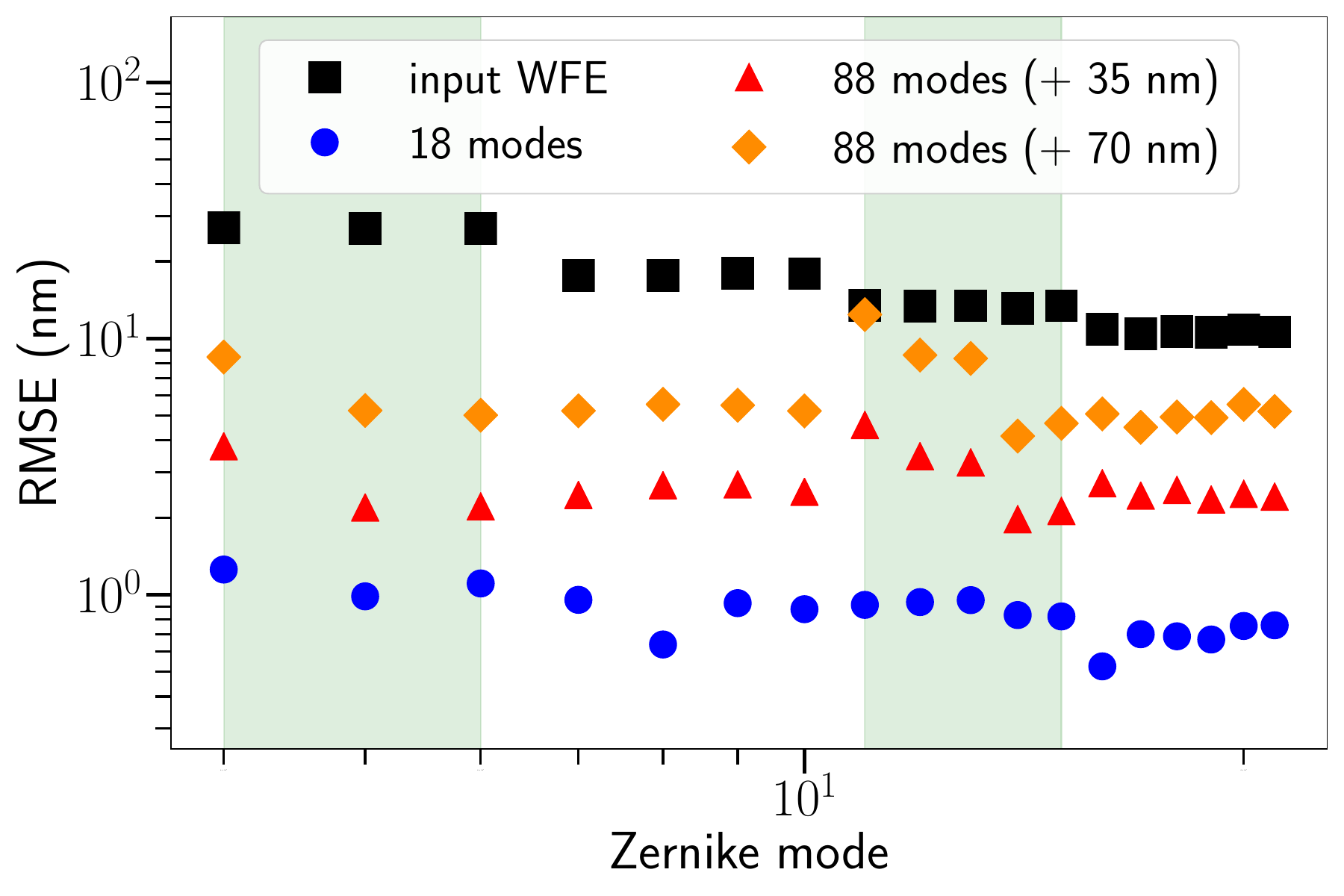}
    \end{subfigure}
    \begin{subfigure}{.99\linewidth}
        \centering
        \includegraphics[width=\hsize]{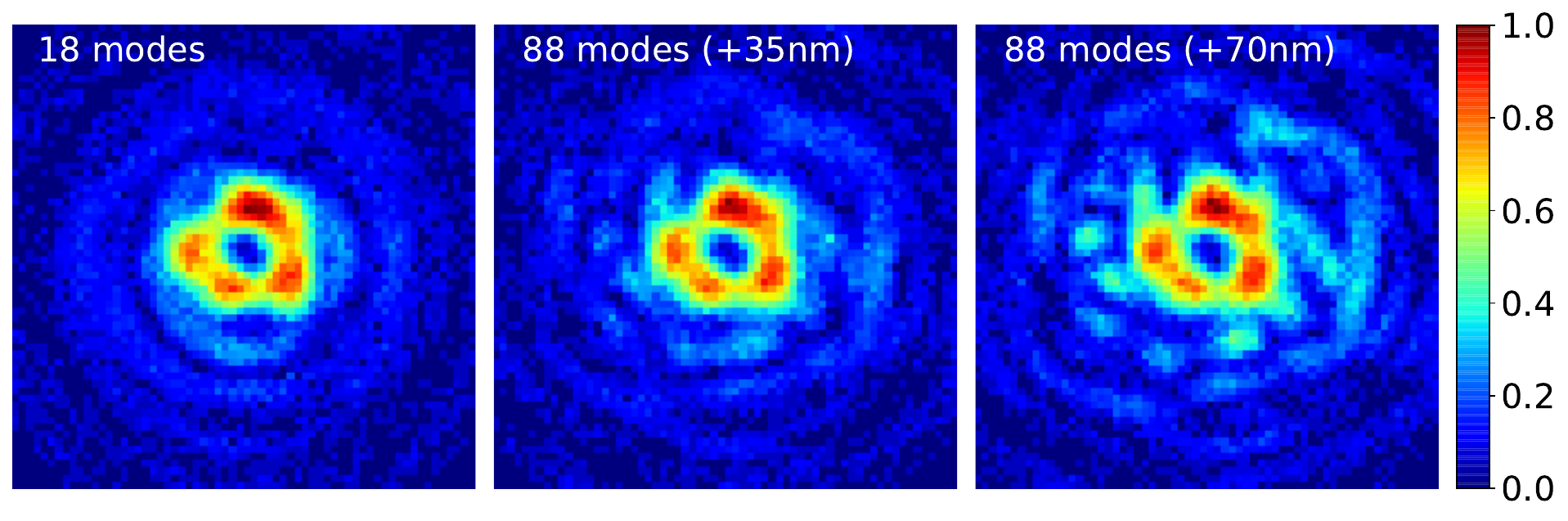}
    \end{subfigure}
    \caption{Robustness on higher-order aberrations. \textit{Top}: Performance per Zernike mode on test data following the distribution of the training data (red), adding, to the test data, 70 modes containing 35~nm RMS (purple) and 70~nm RMS (blue) of NCPAs. \textit{Bottom}: Example of post-VVC PSFs for each case ($+l_p$).}
    \label{fig:robust_zmodes}
\end{figure}

It can also be expected to have observations containing higher-order NCPAs (in addition to the changing atmospheric residuals) than considered during training. For a model trained on 18 modes at 70~nm RMS (S/N = 1000), we added 70 higher-order Zernike modes in the test data. In Fig.~\ref{fig:robust_zmodes}, we observe a moderate degradation for the 18 modes when increasing the wavefront error contained in these additional modes, because the central PSF signature is mostly preserved.

\section{Conclusions}
\label{sec:conclusions}
In this paper, we have investigated a new way to perform focal-plane wavefront sensing using vortex coronagraphs. Based on a deep learning approach and considering simulated data, we have leveraged the modulation introduced by the vortex coronagraph (either scalar, or vectorial after splitting the circular polarization states) to lift the sign ambiguity and perform FPWFS for various S/Ns, input WFEs, and spatial frequency contents. The dual-polarization method with the VVC offers a very similar performance to the classical phase diversity method using additional defocused PSFs, even though the level of light is largely reduced after filtering by the VVC. For instance, considering a star of magnitude 6.2 observed at a wavelength of 2200~nm, we obtain a residual of 0.73~nm RMS from an input WFE of 70~nm RMS. In the case of the SVC, which provides a single focal-plane image, a loss in performance is only observed for high aberration levels. For bright stars, and with higher order and higher levels of aberrations, the CNN training is generally challenging, and the performance reaches a plateau of approximately 20~nm RMS. In such circumstances, more training data, larger and deeper CNN architectures, and regularization techniques could further improve the phase retrieval accuracy.
Atmospheric turbulence residuals that are expected in ground-based data only produce minor degradation in performance in a low NCPA regime, and they should not be a concern in practice. We have also shown that models trained on data containing particularly wide WFE and S/N distributions provide very good robustness.

Potential applications of the proposed method could rely on including a polarizing beam splitter downstream of the VVC to collect both circular polarization states separately, either on a single or on two distinct sensors. Since our simulation-based FPWFS experiments work well even with a single image obtained behind an SVC, it appears that this flavor of vortex coronagraph offers an interesting alternative, notably because it would work without any additional optical components.

Deep learning models offer a flexible framework and fast inference speeds, which are appreciable features for on-sky applications. The requirement on speed is, however, not very stringent as we expect the lifetime of NCPAs that produce quasi-static speckles to be on the order of minutes.
But on-sky applications will naturally come with their own challenges and discrepancies unpredicted by simulations. To account for the difference between simulations and real data, transfer learning techniques can be used to efficiently fine-tune the models before observations. Finally, it is difficult to obtain reliable and very precise NCPA labels for model training. Employing unsupervised learning techniques, for example autoencoder-based architectures, is another interesting approach that we are considering for future developments.

\begin{acknowledgements}
This research made use of PyTorch \citep{Paszke:19} and the following implementation of EfficientNet: \url{https://github.com/lukemelas/EfficientNet-PyTorch}. The HEEPS \citep{Carlomagno:20} and PROPER \citep{Krist:07} open-source optical propagation Python packages were used for data generation. This project has received funding from the European Research Council (ERC) under the European Union's Horizon 2020 research and innovation programme (grant agreement No 819155), and from the Wallonia-Brussels Federation (grant for Concerted Research Actions).
\end{acknowledgements}

%
%

\bibliographystyle{aa}
\bibliography{main}

\begin{thebibliography}{28}
\expandafter\ifx\csname natexlab\endcsname\relax\def\natexlab#1{#1}\fi

\bibitem[{{Absil} {et~al.}(2016){Absil}, {Mawet}, {Karlsson}, {Carlomagno},
  {Christiaens}, {Defr{\`e}re}, {Delacroix}, {Femen{\'\i}a Castella},
  {Forsberg}, {Girard}, {G{\'o}mez Gonz{\'a}lez}, {Habraken}, {Hinz}, {Huby},
  {Jolivet}, {Matthews}, {Milli}, {Orban de Xivry}, {Pantin}, {Piron},
  {Reggiani}, {Ruane}, {Serabyn}, {Surdej}, {Tristram}, {Vargas Catal{\'a}n},
  {Wertz}, \& {Wizinowich}}]{Absil:16}
{Absil}, O., {Mawet}, D., {Karlsson}, M., {et~al.} 2016, in Society of
  Photo-Optical Instrumentation Engineers (SPIE) Conference Series, Vol. 9908,
  Ground-based and Airborne Instrumentation for Astronomy VI, ed. C.~J.
  {Evans}, L.~{Simard}, \& H.~{Takami}, 99080Q

\bibitem[{{Andersen} {et~al.}(2019){Andersen}, {Owner-Petersen}, \&
  {Enmark}}]{Andersen:19}
{Andersen}, T., {Owner-Petersen}, M., \& {Enmark}, A. 2019, Optics Letters, 44,
  4618

\bibitem[{{Andersen} {et~al.}(2020){Andersen}, {Owner-Petersen}, \&
  {Enmark}}]{Andersen:20}
{Andersen}, T., {Owner-Petersen}, M., \& {Enmark}, A. 2020, Journal of
  Astronomical Telescopes, Instruments, and Systems, 6, 034002

\bibitem[{{Bos} {et~al.}(2019){Bos}, {Doelman}, {Lozi}, {Guyon}, {Keller},
  {Miller}, {Jovanovic}, {Martinache}, \& {Snik}}]{Bos:19}
{Bos}, S.~P., {Doelman}, D.~S., {Lozi}, J., {et~al.} 2019, \aap, 632, A48

\bibitem[{{Carlomagno} {et~al.}(2020){Carlomagno}, {Delacroix}, {Absil},
  {Cantalloube}, {Orban de Xivry}, {Pathak}, {Agocs}, {Bertram}, {Brand l},
  {Burtscher}, {Feldt}, {Glauser}, {Hippler}, {Kenworthy}, {Stuik}, \& {van
  Boekel}}]{Carlomagno:20}
{Carlomagno}, B., {Delacroix}, C., {Absil}, O., {et~al.} 2020, Journal of
  Astronomical Telescopes, Instruments, and Systems, 6, 035005

\bibitem[{Desai {et~al.}(2021)Desai, Llop-Sayson, Jovanovic, Ruane, Serabyn,
  Martin, \& Mawet}]{Desai:21}
Desai, N., Llop-Sayson, J., Jovanovic, N., {et~al.} 2021, in Techniques and
  Instrumentation for Detection of Exoplanets X, ed. S.~B. Shaklan \& G.~J.
  Ruane, Vol. 11823, International Society for Optics and Photonics (SPIE), 238
  -- 246

\bibitem[{{Dohlen} {et~al.}(2011){Dohlen}, {Wildi}, {Puget}, {Mouillet}, \&
  {Beuzit}}]{Dohlen:11}
{Dohlen}, K., {Wildi}, F.~P., {Puget}, P., {Mouillet}, D., \& {Beuzit}, J.-L.
  2011, in Second International Conference on Adaptive Optics for Extremely
  Large Telescopes. Online at <A
  href=``http://ao4elt2.lesia.obspm.fr''>http://ao4elt2.lesia.obspm.fr</A, 75

\bibitem[{Ferreira {et~al.}(2018)Ferreira, Gratadour, Sevin, \&
  Doucet}]{Ferreira:18}
Ferreira, F., Gratadour, D., Sevin, A., \& Doucet, N. 2018, 2018 International
  Conference on High Performance Computing \& Simulation (HPCS), 180

\bibitem[{Fienup(1982)}]{Fienup:82}
Fienup, J. 1982, Applied optics, 21, 2758

\bibitem[{Gerchberg(1972)}]{Gerchberg:72}
Gerchberg, R.~W. 1972, Optik, 35, 237

\bibitem[{Gonsalves(1982)}]{Gonsalves:82}
Gonsalves, R.~A. 1982, Optical Engineering, 21, 829

\bibitem[{Guizar-Sicairos \& Fienup(2012)}]{Guizar-Sicairos:12}
Guizar-Sicairos, M. \& Fienup, J.~R. 2012, J. Opt. Soc. Am. A, 29, 2367

\bibitem[{{Guyon}(2018)}]{Guyon:18}
{Guyon}, O. 2018, \araa, 56, 315

\bibitem[{{Jovanovic} {et~al.}(2018){Jovanovic}, {Absil}, {Baudoz}, {Beaulieu},
  {Bottom}, {Cady}, {Carlomagno}, {Carlotti}, {Doelman}, {Fogarty}, {Galicher},
  {Guyon}, {Haffert}, {Huby}, {Jewell}, {Keller}, {Kenworthy}, {Knight},
  {K{\"u}hn}, {Miller}, {Mazoyer}, {N'Diaye}, {Por}, {Pueyo}, {Riggs}, {Ruane},
  {Sirbu}, {Snik}, {Wallace}, {Wilby}, \& {Ygouf}}]{Jovanovic:18}
{Jovanovic}, N., {Absil}, O., {Baudoz}, P., {et~al.} 2018, in Society of
  Photo-Optical Instrumentation Engineers (SPIE) Conference Series, Vol. 10703,
  Adaptive Optics Systems VI, ed. L.~M. {Close}, L.~{Schreiber}, \&
  D.~{Schmidt}, 107031U

\bibitem[{Kingma \& Ba(2017)}]{Kingma:17}
Kingma, D.~P. \& Ba, J. 2017, Adam: A Method for Stochastic Optimization

\bibitem[{Krist(2007)}]{Krist:07}
Krist, J.~E. 2007, in Optical Modeling and Performance Predictions III, ed.
  M.~A. Kahan, Vol. 6675, International Society for Optics and Photonics
  (SPIE), 250 -- 258

\bibitem[{{Lamb} {et~al.}(2021){Lamb}, {Correia}, {Sivanandam}, {Swanson}, \&
  {Zavyalova}}]{Lamb:21}
{Lamb}, M.~P., {Correia}, C., {Sivanandam}, S., {Swanson}, R., \& {Zavyalova},
  P. 2021, \mnras, 505, 3347

\bibitem[{Martinache(2013)}]{Martinache:13}
Martinache, F. 2013, Publications of the Astronomical Society of the Pacific,
  125, 422

\bibitem[{{Mawet} {et~al.}(2005){Mawet}, {Riaud}, {Absil}, \&
  {Surdej}}]{Mawet:05}
{Mawet}, D., {Riaud}, P., {Absil}, O., \& {Surdej}, J. 2005, \apj, 633, 1191

\bibitem[{Mawet {et~al.}(2009)Mawet, Serabyn, Liewer, Burruss, Hickey, \&
  Shemo}]{Mawet:09}
Mawet, D., Serabyn, E., Liewer, K., {et~al.} 2009, The Astrophysical Journal,
  709

\bibitem[{{Orban de Xivry} {et~al.}(2021){Orban de Xivry}, {Quesnel},
  {Vanberg}, {Absil}, \& {Louppe}}]{Orban:21}
{Orban de Xivry}, G., {Quesnel}, M., {Vanberg}, P.~O., {Absil}, O., \&
  {Louppe}, G. 2021, \mnras, 505, 5702

\bibitem[{{Paine} \& {Fienup}(2018)}]{Paine:18}
{Paine}, S.~W. \& {Fienup}, J.~R. 2018, Optics Letters, 43, 1235

\bibitem[{Paszke {et~al.}(2019)Paszke, Gross, Massa, Lerer, Bradbury, Chanan,
  Killeen, Lin, Gimelshein, Antiga, Desmaison, Kopf, Yang, DeVito, Raison,
  Tejani, Chilamkurthy, Steiner, Fang, Bai, \& Chintala}]{Paszke:19}
Paszke, A., Gross, S., Massa, F., {et~al.} 2019, in Advances in Neural
  Information Processing Systems, ed. H.~Wallach, H.~Larochelle,
  A.~Beygelzimer, F.~d\textquotesingle Alch\'{e}-Buc, E.~Fox, \& R.~Garnett,
  Vol.~32 (Curran Associates, Inc.), 8026--8037

\bibitem[{{Quesnel} {et~al.}(2020){Quesnel}, {Orban de Xivry}, {Louppe}, \&
  {Absil}}]{Quesnel:20}
{Quesnel}, M., {Orban de Xivry}, G., {Louppe}, G., \& {Absil}, O. 2020, in
  Society of Photo-Optical Instrumentation Engineers (SPIE) Conference Series,
  Vol. 11448, Society of Photo-Optical Instrumentation Engineers (SPIE)
  Conference Series, 114481G

\bibitem[{{Riaud} {et~al.}(2012{\natexlab{a}}){Riaud}, {Mawet}, \&
  {Magette}}]{Riaud:12b}
{Riaud}, P., {Mawet}, D., \& {Magette}, A. 2012{\natexlab{a}}, \aap, 545, A151

\bibitem[{{Riaud} {et~al.}(2012{\natexlab{b}}){Riaud}, {Mawet}, \&
  {Magette}}]{Riaud:12a}
{Riaud}, P., {Mawet}, D., \& {Magette}, A. 2012{\natexlab{b}}, \aap, 545, A150

\bibitem[{Ruane {et~al.}(2019)Ruane, Mawet, Riggs, \& Serabyn}]{Ruane:19}
Ruane, G., Mawet, D., Riggs, A.~E., \& Serabyn, E. 2019, in Techniques and
  Instrumentation for Detection of Exoplanets IX, ed. S.~B. Shaklan, Vol.
  11117, International Society for Optics and Photonics (SPIE), 454 -- 469

\bibitem[{Tan \& Le(2019)}]{Tan:19}
Tan, M. \& Le, Q. 2019, in Proceedings of Machine Learning Research, Vol.~97,
  Proceedings of the 36th International Conference on Machine Learning, ed.
  K.~Chaudhuri \& R.~Salakhutdinov (PMLR), 6105--6114

\end{thebibliography}

\end{document}